# Direct Exfoliation of Nanoribbons from Bulk van der Waals Crystals

**Ashley P. Saunders[1], Victoria Chen[2], Jierong Wang[3], Amalya C. Johnson[4], Amy S. McKeown-Green[1], Helen J. Zeng[1], T. Kien Mac[6], Tuan Trinh[6], Tony F. Heinz[3,5], Eric Pop[2], Fang Liu[1]\***

[1]Department of Chemistry, Stanford University, Stanford, CA 94305, USA
[2]Department of Electrical Engineering, Stanford University, Stanford, CA 94305, USA
[3]Department of Applied Physics, Stanford University, Stanford, CA 94305, USA
[4]Department of Materials Science and Engineering, Stanford University, Stanford, CA 94305, USA
[5]SLAC National Accelerator Laboratory, Menlo Park, CA 94025, USA
[6]Department of Chemistry and Biochemistry, Utah State University, Logan, UT 84322, USA
*Corresponding author: fliu10@stanford.edu

## Abstract
Confinement of monolayers into quasi-one-dimensional atomically-thin nanoribbons could lead to novel quantum phenomena beyond those achieved in their bulk and monolayer counterparts. However, current experimental availability of nanoribbon species beyond graphene has been limited to bottom-up synthesis or top-down patterning. In this study, we introduce a versatile and direct lithography-free approach to exfoliate a variety of bulk van der Waals (vdW) crystals into nanoribbons. Akin to the Scotch tape exfoliation in producing monolayers, this technique provides convenient access to a wide range of nanoribbons derived from their corresponding bulk crystals, including $MoS_2$, $WS_2$, $MoSe_2$, $WSe_2$, $MoTe_2$, $WTe_2$, $ReS_2$, and hBN. The nanoribbons are single-crystalline, parallel-aligned, flat, and have high aspect ratio. We demonstrated the electrical, magnetic, and optical properties from the confinement, strain, and edge configurations of these nanoribbons. This versatile preparation technique will pave the way for future experimental investigation and broad applications in optoelectronic, sensing, electronic and quantum devices.

## Introduction

Facile and versatile preparation methods are paramount in the development of any emerging class of quantum materials to facilitate fundamental research leading to application advancements. A prime example is the Scotch tape exfoliation technique, which launched the field of two-dimensional (2D) materials.[1] This tape exfoliation acts as a universal top-down strategy by readily separating nearly any bulk van der Waals (vdW) crystal into its monolayer counterparts. It grants easy access to a diverse spectrum of atomically thin materials spanning from insulators to semiconductors and metals, whose experimental availabilities led to numerous breakthroughs in fundamental research of 2D physics, such as topological and ferroelectric properties,[2] correlated electronic behavior,[3–5] or light matter interactions.[6,7] When 2D materials are further engineered into one-dimensional (1D) nanoribbons, the 1D lateral confinement and edge states profoundly influence band structure and govern key electronic transitions.[8,9] As a result, nanoribbons manifest a spectrum of new properties including emergent magnetism, topological phase transitions, spin density waves, and catalytic activity.[10–15]

Nanoribbon materials also offer remarkable versatility in electronic, optoelectronic, magnetic, and thermoelectric applications depending on their width, edge structures, and strain.[10,14,16–23] For instance, the absence of a band gap in graphene limits its use in electronics, while graphene nanoribbons (GNRs) show semiconducting properties accompanied by width-variable band gaps, new topological phases, and localized electronic edge states.[11,18,24–28] Hexagonal boron nitride (hBN) is an insulator with a wide band gap ($E_g \approx 6$ eV), while doping-like edge states will turn hBN nanoribbons into semiconductors or metals.[29–31] Transition metal dichalcogenides (TMDCs), such as $MoS_2$, are semiconductors and nonmagnetic in both bulk and monolayer form. In contrast, first principles computations have predicted that TMDC nanoribbons with zigzag edges are expected to be metallic and ferromagnetic ($MoS_2$ and $WS_2$),[10,22,32–35] while nanoribbons with armchair edges are nonmagnetic and semiconducting.[10,19,22,33]

Despite many theoretical studies on the properties of these quasi-one-dimensional structures, the experimental demonstration of a broad variety of nanoribbons, especially those beyond graphene, has been relatively underexplored. Current techniques for nanoribbon production can be categorized into top-down or bottom-up approaches. Top-down methods primarily use unzipping of nanotubes or lithography and etching techniques to produce striped patterns out of monolayers. Examples include electron beam lithography,[36] dry/wet etching,[37–40] scanning probe lithography,[41–43] direct helium ion beam milling,[44,45] ionic scissoring of macroscopic crystals,[46] or liquid phase etching with a reducing agent through a mechano-chemical process.[47] However, the combination of lithography and etching requires expensive equipment and can result in nanoribbons with rough edges and low crystal quality. On the other hand, bottom-up techniques emphasize directed growth or chemical reactions to synthesize TMDC nanoribbons. Strategies include molecular beam epitaxy,[48,49] guided growth within carbon nanotubes,[50] growth along substrate crystal atomic alignments/steps/edges,[21,51–55] or growth with moving catalyst droplets.[56,57] Additionally, only a limited selection of monolayer nanoribbons from the diverse 2D library has been successfully synthesized to date. Most current methods lack control over size, shape, and adaptability to a broad material selection. The development of a *universal*, scalable, and non-destructive production technique for a broad variety of nanoribbons remains a formidable challenge.

Building upon the advancements in metal-assisted monolayer exfoliation,[58] we introduce a universal, reproducible, and large-scale fabrication technique to achieve parallel-aligned and single-crystalline 2D nanoribbons. The ability to obtain nanoribbons directly from an extensive range of vdW bulk materials will facilitate the exploration and manipulation of the distinct electrical, magnetic, or optical properties in 1D geometry. The results will be instrumental for further breakthroughs that encompass dimensionalities beyond those currently explored in monolayer materials.

# Main

A schematic of our exfoliation technique is illustrated in Fig.1. The slanted surface of an as-grown bulk vdW crystals can exhibit a stepwise pattern with a non-zero polar cut angle, θ, termed the vicinal surface (as depicted in Fig. 1b). To initiate the exfoliation of monolayer nanoribbons, a 100 nm thick layer of gold (Au) is evaporated onto the vicinal surface, followed by a spin-coated polyvinylpyrrolidone (PVP) layer to prevent contamination. When the gold layer is lifted with a rigid and flat thermal release tape (TRT), it effectively exfoliates monolayer steps from the bulk crystal surface in the form of aligned nanoribbons. The TRT/PVP/Au/nanoribbon assembly is subsequently transferred onto a destination substrate, such as SiO$_2$/Si, sapphire, or fused silica. Removal of TRT with heating and dissolving the PVP in water leaves a clean Au layer, which is then etched using KI/I$^-$ solution. This etching process is mild and preserves integrity of many 2D materials.[58,59] After cleaning with water, flat and parallel aligned nanoribbons are obtained on the chosen substrate (Fig. 1c), maintaining the same parallel alignment and single crystallinity of the original bulk crystal. The enhanced yield and high-throughput of this method originates from the substantial binding and adsorption energies of Au on 2D monolayers, which can effectively counteract the interlayer vdW force of a variety of bulk crystals.[58,60,61] Despite the use of Au as an exfoliation medium, X-ray Photoelectron Spectroscopy (XPS) analysis confirms that Au is removed to an undetectable level after etching (see Extended Data Fig. 2), validating cleanliness of the nanoribbon crystal surface.

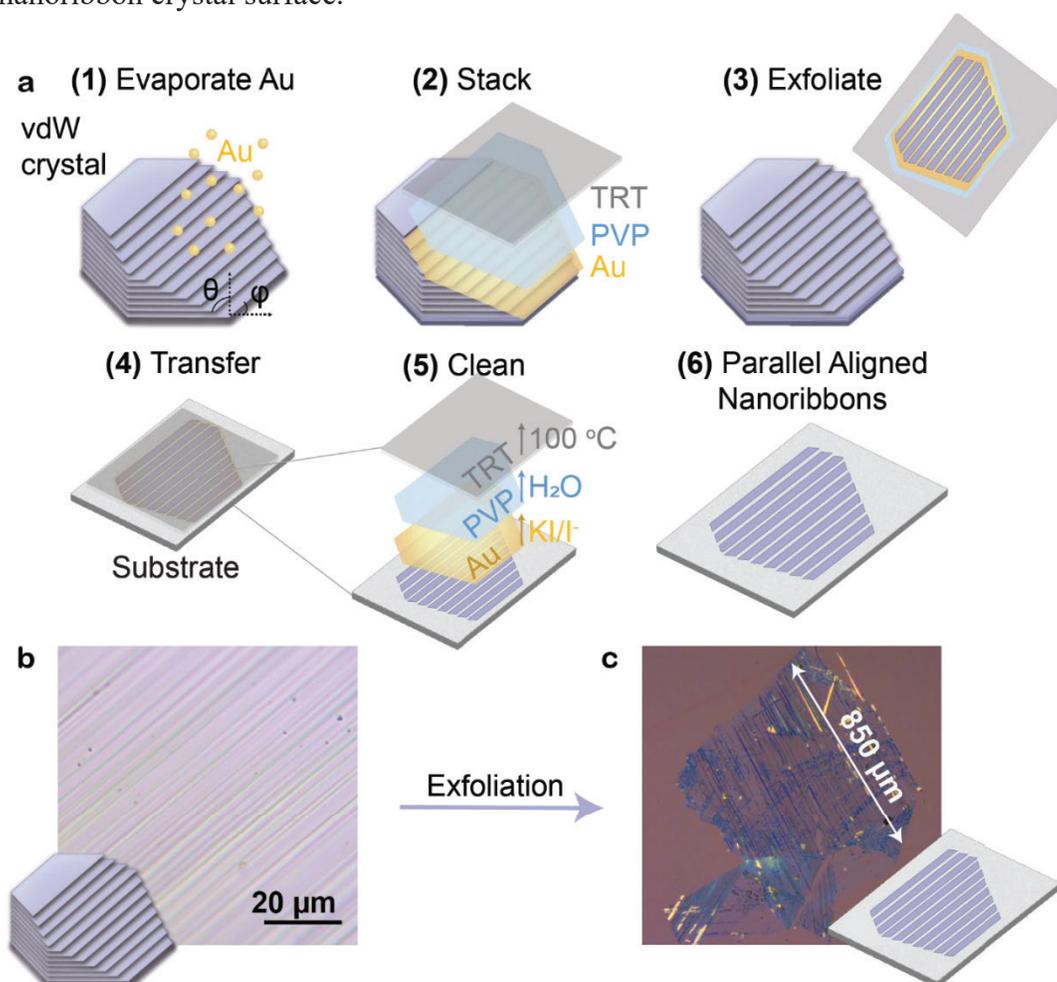

**Fig. 1: Nanoribbon exfoliation from bulk vdW crystals. a**, Schematic illustration of gold-assisted exfoliation technique to obtain single crystalline nanoribbons: (**1**) deposit gold onto a bulk vdW crystal with exposed vicinal edges; (**2**) spin-coat with a layer of PVP and apply TRT; (**3**) pick up the TRT/PVP/Au/nanoribbon stack from the crystal and (**4**) place stack onto the destination substrate; (**5**) remove the TRT by heating at 100 °C, rinse with water to dissolve the sacrificial PVP layer, and etch gold away by submerging in etchant solution; (**6**) obtain parallel aligned single

crystalline nanoribbons. **b**, Optical image of an exemplary vicinal surface on a bulk WSe$_2$ crystal. **c**, Optical image of a large area of continuous MoSe$_2$ nanoribbons on 90 nm SiO$_2$/Si.

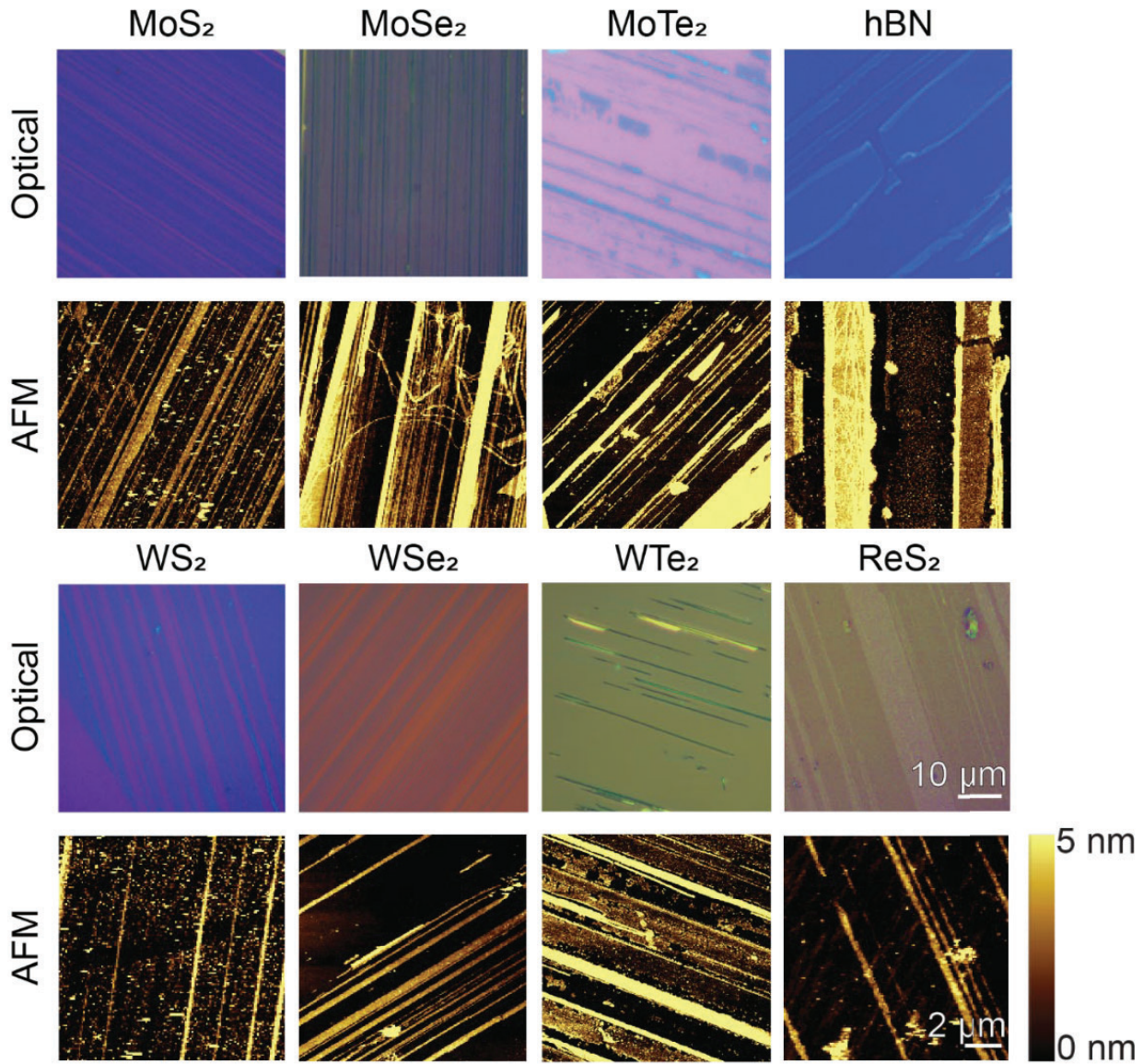

**Fig. 2: Examples of different exfoliated nanoribbons.** Top, Optical images and bottom, atomic force microscopy (AFM) images.

This technique allows us to exfoliate a diverse array of nanoribbons directly from their corresponding vdW bulk crystals. Examples are illustrated in Fig. 2. These nanoribbons encompass a vast spectrum ranging from semiconducting transition metal dichalcogenides to insulating hexagonal boron nitride, and they can be exfoliated onto various destination substrates. Notably, these monolayer nanoribbons preserve the flatness, parallel alignment, straight edges, and single-crystalline nature of their initial bulk crystal surfaces. These nanoribbons feature very large aspect ratios, including widths varying from 20 nm to several hundreds of nanometers, and lengths extending up to the millimeter scale. The dimensions of these ribbons, incorporating their length, width and spacing, directly correlate with the length, width, and height of original crystal steps present on the vicinal surface (Fig. 1b). The efficiency in production of parallel-aligned nanoribbons over hundreds of micrometers is due to the robust and universal interaction between gold and 2D monolayers. This stands in contrast to existing fabrication and synthesis techniques, which often face challenges in large scale nanoribbon production, transferring ribbons from their growth substrates for device integration, and/or compatibility with a diverse range of 2D materials. Our technique provides an important steppingstone for further experimental characterizations on a broad variety of nanoribbons.

TMDC nanoribbons exhibit photoluminescence (PL) emissions that cover a spectrum from the visible to near-infrared range, comparable to their monolayer counterparts. Figs. 3a and 3b compare the PL emissions from exfoliated $WSe_2$ nanoribbons of widths varying from hundreds of nanometers down to tens of nanometers. A notable trend can be observed, where narrower nanoribbons emit at higher energies than wider ones. To establish a direct comparison with TMDC monolayers, we employed the gold-assisted exfoliation technique to obtain monolayers and nanoribbons from the same vdW bulk crystal. As shown in Fig. 3b, despite the distribution of PL peak position over different ribbon widths, nanoribbon PL peak positions are still located within the wide distribution of PL emission of their monolayer equivalents, albeit shifted slightly towards the higher energy end. Previous studies on $MoS_2$ nanoribbons synthesized on Si(001) surfaces pretreated with phosphine, or through controlled $O_2$ etching of monolayers observe PL blueshifts up to 70 meV as the nanoribbon width is reduced below 50 nm.[39,51] Joint with our own study, this same trend of PL blueshift in narrower nanoribbon widths is consistent amongst differing production methods. However, this trend opposes the anticipated redshift that is often induced by defects, while the width of ribbons (tens of nm) is greater than the exciton radius or length in the quantum confinement regime.

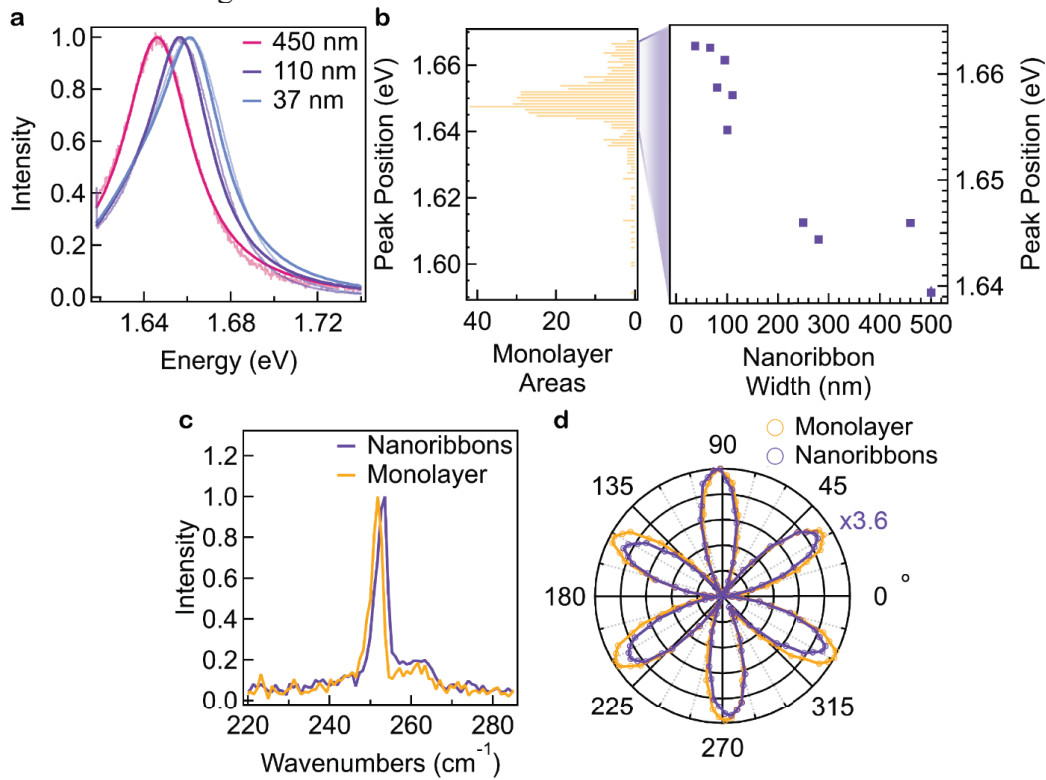

**Fig. 3: Optical spectroscopy of WSe$_2$ nanoribbons. a**, Photoluminescence spectra. The magenta, purple, and blue spectra correspond to nanoribbons of 450 nm, 110 nm, and 37 nm widths respectively. Widths were identified via AFM. The solid lines are Lorentzian fit functions. **b**, **left**, Distribution of PL emission peak wavelength for exfoliated WSe$_2$ monolayers; **right**, PL peak position of WSe$_2$ nanoribbons with different widths, exfoliated from the same crystal. Each point is the averaged signal from several spots along a ribbon spaced over at least 1 μm. **c**, Raman spectra for an area of nanoribbons (purple) compared with an unstrained monolayer area (gold). **d**, Normalized second harmonic generation polarization scans for WSe$_2$ nanoribbons (purple) in comparison to a WSe$_2$ monolayer (gold).

A possible mechanism for this blueshift is compressive strain.[62,63] As shown in Fig. 3d, the E′ Raman mode for WSe$_2$ nanoribbons (253.1 cm$^{-1}$) resides at a higher frequency than the same mode in monolayer flakes (252.7 cm$^{-1}$). This shift is consistent among multiple locations of WSe$_2$ monolayer and nanoribbons (Extended Data Fig. 3). In TMDC monolayers, a shift of the E′ Raman peak by 0.46 cm$^{-1}$ to higher frequency correlates to a compressive strain of 1.62% on average, as estimated in strained WSe$_2$ monolayers.[64] The impact of strain within these exfoliated nanoribbons also affects their second harmonic generation (SHG). By scanning the pump light polarization relative to crystal orientation, unstrained monolayers typically exhibit symmetrical SHG polarization response, signifying the rotational symmetry of their hexagonal crystal lattices (Fig. 3d). In comparison, nanoribbons exhibit a slightly asymmetrical SHG signal, suggesting intrinsic compressive lattice strain-oriented perpendicular to the ribbon direction (Extended Data Fig. 4). Strain appears to be particularly pronounced as a 2D lattice is reduced into 1D structures, indicating that narrow ribbons can be more susceptible to mechanical deformation.

Many vdW bulk crystals grow with zigzag edge termination due to its chemical stability over the armchair configuration.[65] As a result, our exfoliated nanoribbons are predominantly oriented along the zigzag direction. The outer edge of the ribbons maintain the zigzag geometry of bulk crystal edges, while the inner edge is created from guided tearing where fracture of the monolayer follows the direction of the straight crystal zigzag step on top.[66] The crystal orientation of these nanoribbons was verified with SHG polarization scans (Fig. 3d). First principles computations predict that MoS$_2$ and WS$_2$ nanoribbons with zigzag edges are ferromagnetic and metallic,[10,22,32,33] with a magnetic moment that increases as width increases so long as the ribbon remains at the nano-scale.[10] Previous scanning tunneling microscopy (STM) investigations of zigzag edged TMDC nanoribbons showed that wider ribbons change from metallic at the edge to semiconducting at the center of the ribbon.[21] Large TMDC flakes and bulk crystals have also been reported to exhibit weak ferromagnetic behavior,[67–71] which is enhanced for single layers and small grain sizes.[68,70,72] However, the ferromagnetic responses of TMDC bulk crystals and monolayers can be influenced by multiple factors, such as metal vacancies,[73] chalcogen-metal antisites,[72,74] crystal edges, or grain boundaries[34,68,69,72,73]. The interplay of these factors could significantly impact the magnetic properties of these materials.

To identify if exfoliated nanoribbons with zigzag edges are ferromagnetic, we performed magnetic force microscopy (MFM) measurements MoS$_2$ nanoribbons. This two-pass AFM technique records topography and corresponding cantilever oscillation amplitude and phase at a lifted height, providing insight into the magnetic or electrostatic interaction forces exerted on the magnetic AFM probe.[75–78] Figure 4 shows the MFM oscillation phase images of MoS$_2$ nanoribbons and larger area MoS$_2$ 2D flakes exfoliated from the same bulk crystal and processed under the same chemical environment during fabrication and imaging. The MFM phase shift, calculated by subtracting the response from the substrate, appears uniform across each nanoribbon as it reaches the limit of spatial resolution (Extended Data Fig. 5). The edges on a 2D flake were slightly repulsive to the AFM probe, with a very small positive phase shift (Fig. 4c). In contrast, for nanoribbons of various widths less than 300 nm, the phase shift of lifted cantilever oscillation is consistently more negative and therefore attractive (Fig. 4d). A histogram demonstrating the phase shifts from multiple samples is shown in Fig. 4k. The negative phase is consistent among multiple ribbon samples, with wider nanoribbons exhibiting slightly higher attraction than narrower ribbons.

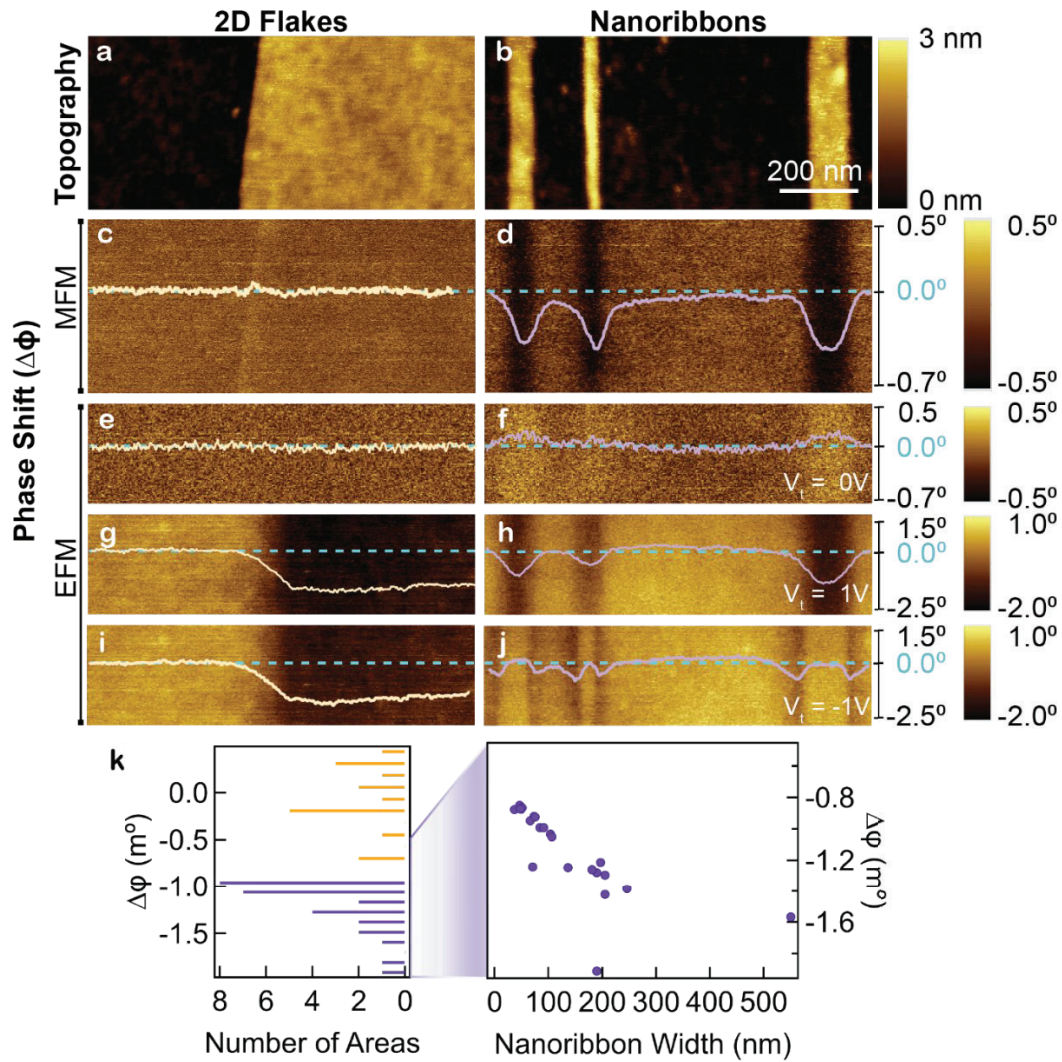

**Fig. 4: Scanning Probe Microscopy of 2D MoS$_2$ flakes and MoS$_2$ nanoribbons. a**, **b**, AFM topography images. **c**, **d**, Magnetic force microscopy oscillation phase images, with an overlayed profile of the MFM oscillation phase response and a zero-degree line (blue) for reference. **e – h**, Electrostatic force microscopy oscillation phase image at different tip voltages (V$_t$) (**e**, **f**) V$_t$ = 0 V, (**g**, **h**) V$_t$ = +1 V, and (**i**, **j**) V$_t$ = –1V with an overlayed profile of the EFM oscillation phase response. **k**, **left**, Histogram comparing the MFM phase shift values among nanoribbons (purple)

and edges of 2D flakes (gold) of similar thickness with (**right**) MFM oscillation phase of nanoribbons of differing widths. The 2D flakes and nanoribbons are exfoliated from the same crystal with similar thickness. All MFM/EFM measurements were carried out at 15 nm lift height, and phase traces subtract the phase response of the substrate.

To comprehensively assess the magnetic behavior, we conducted both reflective magnetic circular dichroism (RMCD), a microscope-based spatial sensitive technique, and vibrating sample magnetometry (VSM), a macroscopic characterization technique (Extended Data Fig. 7 – 8). We do not observe ferromagnetism within individual exfoliated nanoribbons, however, very weak ferromagnetic behaviors are detected in densely packed nanoribbons and monolayers on a large scale. Indeed, when nanoribbons are grounded (Fig. 4c), the attractive MFM phase disappears, indicating that the underlying mechanism of the attractive MFM response is an electrostatic interaction instead of a magnetic one. The presence of static charges may have arisen from uneven chemical bond cleavage as the ribbons are torn from the bulk crystal during exfoliation process.

To illustrate the microscopic electrostatic interaction, we conducted electrostatic force microscopy (EFM) on nanoribbons and 2D flakes exfoliated from the same crystal, using the same probe and lift height. In an EFM measurement, the tip oscillates above the sample while it is electrically biased. The measured phase shift of the tip reflects the electrostatic force gradient, which is related to the potential difference between the tip and the sample.[79] When a positive ($V_t = +1$ V) or negative ($V_t = -1$ V) bias is applied to the tip (Fig. 4g-j), the EFM phase of the 2D flake is consistently negative. Similarly, the EFM phase in nanoribbons is negative at $V_t = +1$ V. However, when the bias is reversed ($V_t = -1$ V), the response dissipates at the center of the nanoribbons, while the negative phase response becomes more pronounced at the nanoribbon edges (Fig. 4h and 4j). The different phase shifts under negative bias indicate that nanoribbon edges may have more *p*-doped character, which is consistent with the $MoS_2$ nanoribbons terminated with bare sulfur zigzag edges.[80] The nanoribbon body is more likely doped *n*-type, consistent with what is commonly found in TMDC monolayers with S vacancies. Similar characteristics can be observed in the amplitude signal. (Extended data Fig. 6)

We also probed the electrical properties of exfoliated nanoribbons through their performance in field-effect transistors. Fabricated devices and measurements of exfoliated, parallel aligned multilayer $MoS_2$ nanoribbons with Au electrodes are displayed in Fig. 5. (See Materials and Methods for fabrication and measurement details.) Each device comprises several nanoribbons and multiple devices were measured with varying average nanoribbon widths ($W_{avg}$). The current density ($I_D$) through the nanoribbon devices was normalized by the totaled width of the nanoribbons. As the average ribbon width narrows, the minimum current measured ($I_{min}$) increases while the ratio of maximum to minimum current ($I_{max}/I_{min}$) decreases, as shown in Fig. 5b and 5c. This indicates that the nanoribbon edges are partially conductive, increasing the minimum current through their zigzag (metallic)[10,32–35] character or by introducing defect states in the band gap. When the devices are turned on, the edge roughness introduces additional scattering, decreasing the maximum current.

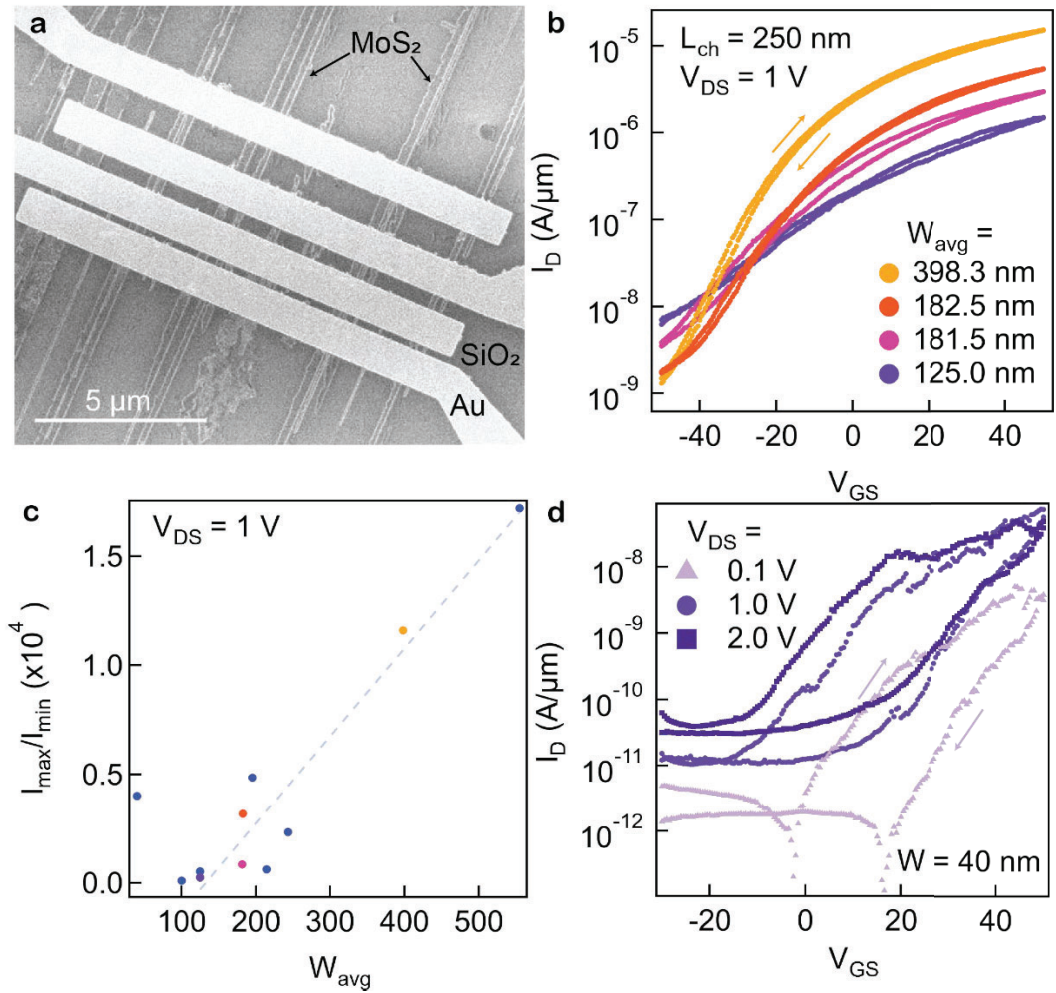

**Fig. 5: Electronic transport properties of MoS$_2$ nanoribbons. a**, Top-down, scanning electron microscopy (SEM) image of parallel MoS$_2$ nanoribbons with Au contacts. The devices are on SiO$_2$ (90 nm) back-gated by the highly doped Si below. **b**, Transfer characteristics of parallel nanoribbons with different average ribbon width ($W_{avg}$) between electrodes. Two curves of each kind correspond to forward and backward $V_{GS}$ sweeps, revealing little hysteresis. **c**, On/off current ratio for several devices with different average ribbon widths. The devices have a channel length (electrode separation) of 250 nm and $V_{DS}$ = 1.0 V between the Au electrodes. **d**, Measured transfer characteristics of an isolated 40 nm wide nanoribbon, with different $V_{DS}$. Dual sweeps marked with small arrows represent forward and backward $V_{GS}$ sweeps.

The $I_{max}/I_{min}$ ratio for these nanoribbons reaches up to 10$^4$, which is comparable with that previously reported for MoS$_2$ monolayer nanoribbons created via lithography and select chemical vapor deposition growth,[37,41] as well as nanoribbons of bulk thickness created through plasma dry-etching and collapsed nanotubes[37,81]. Threshold voltages and field-effect mobilities can be estimated from the transport measurements (Extended Data Fig. 9). The mobility follows a similar trend where narrower average nanoribbon widths generally display lower mobilities. When a single narrow nanoribbon is isolated between two Au electrodes (Fig. 5d), its transfer characteristics display greater hysteresis between the forward and backwards $V_{GS}$ sweeps, with a slight reduction in the $I_{max}/I_{min}$ ratio. This could also be due to an increased influence of mid-gap edge defects which act as charge trapping centers. Compared to much wider monolayer devices,[82] the pronounced influence of edge properties in nanoribbons enhances their potential for applications in chemical sensing, electrocatalysis, and photocatalysis. These applications encompass a range of processes, including but not limited to water splitting, hydrodesulfurization, and $CO_2$ reduction. The successful preparation of nanoribbons with parallel-aligned edges at high-density and their integration into devices suggests promising directions for future research in chemical sensing and opens up opportunities to explore and improve electrocatalytic reaction mechanisms.

## Conclusion

In this work, we introduced a versatile top-down exfoliation approach to obtain single crystal 2D nanoribbons, directly from their corresponding bulk vdW crystals, reminiscent of the renowned Scotch tape exfoliation used for monolayer preparation. The high-throughput approach facilitates the production of high-density, flat, and parallel-aligned nanoribbons from a diverse range of materials, covering large areas on various target substrates. Such capability is crucial for the extensive exploration of the nanoribbons' distinct attributes through spectroscopic, imaging, and transport techniques. This top-down exfoliation method also offers a possibility for versatile control of the nanoribbon morphology. Specifically, the nanoribbon width is determined by the polar cut angle θ, and the atomic alignment along the edge of the nanoribbons is determined by the azimuthal angle φ along the XY plane. If atomically flat crystal planes can be prepared with precisely controlled slant angles, nanoribbons with specifically tailored edge morphology and width can be realized. Examples of such customization include TMDC nanoribbons with semiconducting armchair edges, conducting zigzag edges, or with particular cut directions in between the two geometries. This newfound accessibility to a broad variety of nanoribbon samples will unlock empirically observed properties expanding upon existing theoretical frameworks. It holds the potential to significantly contribute to a range of future technological applications for various nanoribbons, bridging the gap between theoretical predictions and practical implementation.

# Materials and Methods

## Nanoribbon Exfoliation

TMDC crystals (HQ Graphene) and hBN crystals (2D Semiconductors) with slanted edges are deposited with a 100 nm thick layer of Au with an ebeam evaporator (Kurt J. Lesker LAB18), at deposition rate of 0.5 Å/s. The crystals coated with Au are then spin coated with at 10 wt% polyvinyl pyrrolidone (Alfa Aesar) in 1:1 ethanol (Fisher Chemical) to acetonitrile (Fisher Chemical, >99.95%) at 400 rpm for two minutes. Next, a piece of thermal release tape (TRT, Semiconductor Equipment Corp., 90 °C release) is applied to a crystal area covered in Au and PVP with exposed vicinal edges. The TRT is gently lifted, whilst keeping it flat, to remove the Au and top layer/layers of the crystal. The TRT/PVP/Au/nanoribbon stack is then placed, nanoribbon side down, onto the desired substrate (90 nm or 285 nm $SiO_2$ on Si, Nova Wafers). The TRT/PVP/Au/nanoribbon/substrate stack is placed onto a hot plate at 100 °C to release the TRT. The PVP is then rinsed away thoroughly in DI water. The Au/nanoribbon/substrate stacks are then placed into a $KI/I^-$ gold etchant (2.2 wt% $I_2$ from Spectrum Chemical and 8.9 wt% KI from Baker Chemical in DI water) for at least 4 minutes. Ultimately, samples are rinsed in DI water, then isopropanol (Fisher Chemical, >99.5), and promptly dried with a $N_2$ gun.

## Photoluminescence and Raman Spectroscopy

Photoluminescence and Raman spectra were taken under room temperature inside a home-built nitrogen cell using a home-built optical spectroscopy setup. A 532 nm cw laser (CNI laser, at 33 µW for PL and 110 µW for Raman) was focused on the nanoribbons to the diffraction limit with a 150 X objective (Nikon Confocal Plan Apochromat). The emitted light from the nanoribbon passes through a 550 nm longpass filter (Thorlabs) for PL and volume Bragg gratings (opt iGRATE) for Raman spectroscopy measurements. The photoluminescence and Raman spectra were collected via a spectrograph (Princeton Instruments, HRS-300-S) and a camera (Princeton Instruments, PIX-400BR). To extract central wavelength, peaks in the recorded spectra were fit with Lorentzian functions.

## Second Harmonic Generation

Second Harmonic Generation was achieved at room temperature using a 1030 nm pulse generated from a femtosecond laser (NKT Origami Onefive 10, pulse duration < 200 fs). This pump pulse was then reflected off an 820 nm short-pass dichroic mirror and passed through a rotating half waveplate before passing through a 40 X Nikon Plan Fluor objective. The objective focusses the beam on the sample whilst under $N_2$. The reflected beam is then directed back up into the objective, through the rotated half wave plate, transmitted through the 820 nm short pass dichroic mirror, and passed through a 600 nm short pass filter. The remaining signal is then collected and amplified by an EM CCD (Andor iXon Ultra), with the simultaneous polarization rotation in both the incoming and outgoing beams resulting in an SHG response with 6-fold symmetry.

**Reflection Magnetic Circular Dichroism**

The existence of magnetic moment inside material can result in different reflectivity for left-circularly polarized (LCP) light and right-circularly polarized (RCP) light, with the difference (RMCD percentage) approximately proportional to the magnetic moment.[71] Therefore magnetic materials can cause incident linearly polarized light to become elliptically polarized reflected light, with ellipticity being a measure of RMCD percentage.[83]

RMCD signal was measured with the sample loaded in Attocube attoDRY2100 cryostat, which can apply up to 9 T magnetic field with an optical path design closely following Huang, et. al[71] and Sato[83]. A continuous wave stabilized 532 nm laser (Coherent Verdi V5) was chopped at 137 Hz with linear polarization at 45° to the fast axis of photo-elastic modulator (PEM, Hind Instruments, PEM series I). The PEM had 50 kHz sinusoidal retardation, with maximum retardance of $\lambda/4$ ($\lambda$ is 532 nm here). The linearly polarized light, passing PEM, was then reflected from the sample, and collected by photodiode (Thorlabs, DET36A). Lock-ins were used to separately obtain signals at the chopper frequency and PEM frequency, from which the reflectivity difference (RMCD percentage) was estimated.[83]

**Atomic Force Microscopy**

Topography images of nanoribbon and monolayer samples were collected at room temperature with Atomic Force Microscopy (Park Systems, Park NX-10, Park XE-70, or XE-100). All Magnetic Force Microscopy and Electrostatic Force Microscopy measurements were taken using a Park NX-10 and a Nanosensors PPP-LC-MFMR probe at a 15 nm lift height above the sample surface. Topography and MFM/EFM phase profiles were obtained from averaging line scans over 32 nm of a 1 x 1 um images.

**Electrical Measurements**

To characterize the electronic transport properties, field-effect transistors were fabricated on 90 nm $SiO_2$ on $p^{++}$ Si substrates, which also served as the back-gate. The larger contact pads and leads were patterned with electron-beam (e-beam) lithography using PMMA (polymethyl methacrylate) as the resist and consist of an e-beam evaporated stack with 25 nm $SiO_2$, 3 nm Ti, and 40 nm Au. (The $SiO_2$ layer reduces the leakage from the large contact pad to the substrate and the Ti layer improves the adhesion of the Au.) The finer leads, which make direct contact with the $MoS_2$, were similarly patterned with e-beam lithography and consist of 40 nm of pure Au deposited with e-beam evaporation at low pressure ($10^{-7}$ Torr or lower).[84] Lift-off of the contacts and pads was done in room temperature acetone. These devices were measured under vacuum ($<10^{-5}$ Torr) at room temperature in a Janis ST-100 probe station with a Keithley 4200 Semiconductor Characterization System.

**Acknowledgements** F.L acknowledges support of the Terman Fellowship and startup grant from the Stanford University Department of Chemistry. The preparation of nanoribbon materials is based upon work supported by the Defense Advanced Research Projects Agency (DARPA) under Agreement No. HR00112390108. A.C.J. acknowledges support from the DOE Office of Science Graduate Student Research (SCGSR) award and TomKat Center Graduate Fellowship for Translational Research. A.M.G. acknowledges support of the National Science Foundation Graduate Research fellowship (NSF-GRFP) under Grant No. DGE-2146755 and the John Stauffer Graduate Fellowship. V.C. and E.P. acknowledge partial support from the Stanford SystemX Alliance and from SUPREME, a Semiconductor Research Corporation (SRC) Center co-sponsored by DARPA. Part of this work was performed at the Stanford Nano Shared Facilities (SNSF), supported by the National Science Foundation under award ECCS-2026822.

**Author Contributions** F.L. conceived the work. A.P.S. performed Au-assisted exfoliations, and all optical imaging and AFM based characterizations apart from the exfoliation of hBN (performed by A.M.G.), the exfoliation of $MoTe_2$ (performed by H.J.Z), and the exfoliation and topographical image of $ReS_2$ (performed by A.C.J). A.P.S. carried out all photoluminescence spectroscopy, Raman spectroscopy, and second harmonic generation characterizations. V.C. performed device fabrication and electrical transport analysis, with advice from E.P. J.W. collected RMCD and aided A.P.S. in analysis of results. T.T. and T.K.M performed all VSM measurements and analysis. A.P.S. and F.L. wrote this manuscript with input from all authors, and all authors contributed to the discussion and interpretation of the data presented in this work.

**Competing Interests** The authors declare no competing interests.

**Materials & Correspondence** Correspondence and requests for materials should be addressed to Fang Liu at fliu10@stanford.edu.